# All van der Waals Integrated Nanophotonics

Haonan Ling[1], Renjie Li[1] and Artur R. Davoyan[1*]

[1]Department of Mechanical and Aerospace Engineering, University of California, Los Angeles, CA, 90095 USA

davoyan@seas.ucla.edu

*Abstract*

Integrated optics is at the heart of a wide range of systems from remote sensing and communications to computing and quantum information processing. Demand for smaller and more energy efficient structures stimulates search for more advanced material platforms. Here, we propose a concept of an all van der Waals photonics, where we show that electronically bulk transition metal dichalcogenide (TMDC) semiconductors are well fitted for the design of key optical components for nanoscale and integrated photonics. Specifically, we demonstrate theoretically that owing to low optical loss and high refractive index across near-infrared and telecom frequency bands, components made of bulk TMDCs can potentially outperform counterparts made of conventional 3D semiconductors, such as Si and III/Vs. We discuss several key quantum and classical optical components and show that bulk TMDCs may pave the way to smaller footprint devices, more energy efficient electro-optical modulators, and stronger quantum light-materials interaction. Enhanced optical performance, ease of integration, and a wide selection of materials suggest that bulk TMDCs may complement and, potentially, replace existing integrated photonics systems.

*Main text*

Transition metal dichalcogenides are an emerging class of layered two-dimensional semiconductors that are of great interest for photonics owing to their unique optical and electronic properties [1,2]. Unlike covalent materials, TMDCs are comprised of weakly bonded atomically thin layers, which can be isolated into stable, free-standing monolayer films [1-3]. A range of notable phenomena unique to monolayer TMDC materials have been discovered [1-4], promising a span of intriguing applications from excitonic light sources [5-8] to atomically thin photovoltaics [9-11]. Integration of 2D materials with conventional optical components and circuits suggests efficient pathways for optical modulation [12-14] and photodetection [15-18]. In these systems light is guided and controlled *within* regular materials (e.g., inside a silicon waveguide); TMDCs interact with optical fields via evanescent coupling (e.g., by placing monolayer TMDC atop of a waveguide). However, the operation limits of integrated optical devices and systems – that is, size, power consumption, classical and quantum light-material interactions – are naturally constrained by the properties of underlying materials constituting the core of integrated photonic circuitry. Especially, material refractive index directly relates to density of optical states and electromagnetic field confinement, which, in turn, translates onto the strength of light-materials interaction, component footprint, and functionality of integrated optoelectronic circuits.

Bulk TMDCs (i.e., electronically thick) possess naturally high refractive indices [19-21]. In particular, $n > 4$ for tungsten and molybdenum sulfides and selenides (i.e., $MoS_2$, $MoSe_2$, $WS_2$, $WSe_2$) in near infrared and telecom frequency bands, which is about 25% higher than that for silicon (Si) and III/Vs (e.g., GaA and InP) widely used in today's integrated photonics industry. Nonetheless, unlike monolayers, bulk TMDCs have received little attention [21-26], with a major focus on enhanced exciton-polariton coupling near resonant exciton absorption peak [21-24].

Here, we propose a concept of an *all van der Waals photonics*, in which light is guided, manipulated and processed at telecom frequencies ($\lambda \geq 1000nm$) within optical components made of electronically bulk TMDCs (i.e., $\geq 5nm$ thick). We show that very high refractive index manifests in strong light-materials interaction and in deeply subwavelength optical field confinement. In contrast to previous works focusing on excitonic regimes of light-materials interaction [21-24], we study light guiding and control below the bandgap, where optical attenuation is small. We demonstrate that optical components made of TMDC materials may outperform conventional optoelectronic semiconductors, such as Si and III/Vs, paving the way to devices with a significantly reduced footprint, strongly enhanced quantum light-materials interaction, and more power-efficient electro-optical switching. Our analysis suggests that TMDCs may complement or replace traditional semiconductors in the design of nanoscale integrated photonic systems.

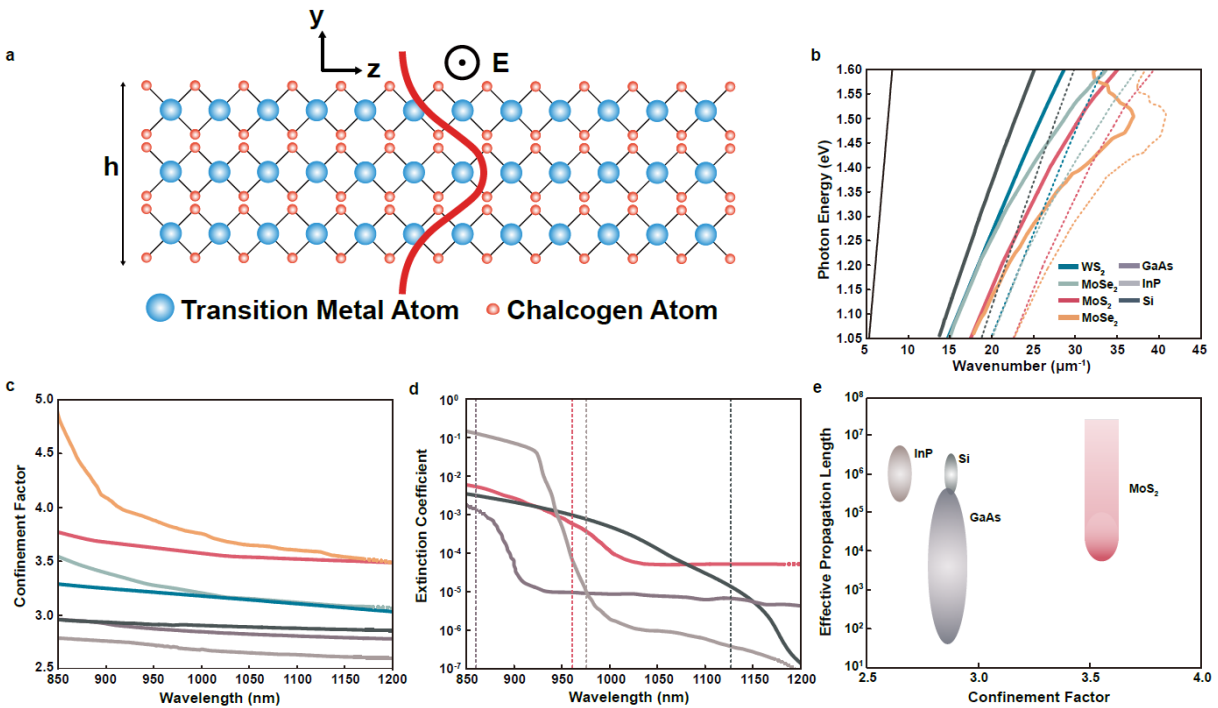

FIG. 1. Performance of planar TMDC waveguides. (a) Schematic illustration of a planar 1D waveguide made of a TMDC material. (b) Dispersion diagram of guided modes in waveguides made of WS$_2$, WSe$_2$, MoS$_2$, MoSe$_2$ and Si (only fundamental TE modes are shown). Dashed curves denote respective material light-lines. (c) Optimal confinement factors for waveguides made of TMDCs and conventional semiconductors (i.e., Si, InP, and GaAs). (d) Extinction coefficients of MoS$_2$, Si, GaAs and InP per ellipsometry and photothermal deflection spectroscopy [27, 29, 34, 35]. Vertical dashed lines represent onset of corresponding material bandgap absorption edges. (e) Performance comparison for MoS$_2$, Si, GaAs, and InP, respectively. Bubble size indicates variation of parameters in a wavelength range from badgap edge to 1200 nm for respective materials.

To illustrate promise of bulk TMDC materials for integrated optics, we begin our analysis with a study of light guiding in a simple one-dimensional planar waveguide, shown in Fig. 1a.

Specifically, we consider electronically bulk (i.e., materials thick enough so that their electronic properties are unperturbed [1], $h \geq 5nm$ ) chalcogenides of tungsten and molybdenum: tungsten disulfide (WS$_2$), tungsten diselenide (WSe$_2$), molybdenum disulfide (MoS$_2$), and molybdenum diselenide (MoSe$_2$). These materials are layered van der Waals semiconductors possessing high refractive index, $n$, in the near infrared wavelength range. In particular, the refractive index along basal planes for these materials ranges from $n_\perp \simeq 3.9$ for tungsten compounds to $n_\perp \simeq 4.4$ for molybdenum compounds [19-21] for $\lambda \geq 1000\ nm$, i.e., below the electronic bandgap ($\hbar\omega \leq 1.2\ eV$) where optical absorbance is low [27,28] (cf. to $n_{Si} \simeq 3.5$ for Si [29] in the same wavelength range). While there are materials with even higher refractive indices, such as germanium (Ge), typically they possess a relatively small electronic bandgap ($\simeq 0.8$ eV for Ge [30]) that limits their use for telecom applications. We also stress that bulk TMDCs have been utilized for design of ultrathin metasurfaces [25] and structures relying on light-exciton interaction [21,23,24]. However, a detailed study of light manipulation within bulk TMDC and their use for integrated optics is yet to be performed.

For the sake of concept demonstration, we study TE polarized waves with the electric field component perpendicular to TMDC crystallographic axis (Fig. 1a). In this case, only in-plane refractive index, $n_\perp$, contributes to the wave dispersion, despite strong natural birefringence of van der Waals materials [19-21, 26]. We account for the full tensorial nature of optical properties in our subsequent discussion. Fig. 1b shows an electromagnetic wave dispersion for a $h = 120nm$ thick waveguide made of TMDC materials of interest (i.e., MoS$_2$, MoSe$_2$, WSe$_2$, and WS$_2$). Dispersion of an equivalent Si waveguide is plotted for comparison. We find that for any given photon energy, $\hbar\omega$, the wavenumber, $k$, for TMDC waveguides is much larger than that of a Si counterpart, indicating a shorter effective guided wavelength, $\lambda_{eff} = \frac{2\pi}{k} \ll \lambda_0$. Hence, effective wavelength in MoS$_2$ waveguide at $\lambda = 1200\ nm$ (i.e., $\hbar\omega \simeq 1.03$ eV) is almost four times smaller than the free space wavelength ($\lambda_{eff} \simeq 350$ nm). Molybdenum compounds, owing to a higher index of refraction as compared to tungsten compounds [19-21], possess larger wavenumbers and, therefore, shorter effective guided wavelengths. Furthermore, for frequencies, $\hbar\omega \geq 1.15\ eV$, (i.e., $\lambda \leq 1070nm$), molybdenum compounds are capable of guiding optical modes that are not supported by silicon (which is manifested as crossing of Si light line by MoS$_2$ and MoSe$_2$ guided wave branches, i.e., $k_{TMDC} > \frac{2\pi}{\lambda} n_{Si}$). The latter opens a possibility for using silicon as a substrate for TMDC waveguides and optical circuits with a minimal field leakage into the substrate [31].

Large wavenumber, $k \gg k_0$, and short effective wavelength, $\lambda_{eff} \ll \lambda$, suggest strong optical field localization, resulting in a small effective mode area for TMDC waveguides. To examine field localization, we introduce an optimal optical confinement, $C_{opt}$, and study its wavelength dispersion. Specifically, we define $C_{opt}(\lambda) = \max_h A_o/A_{eff}$, where $A_o = \frac{\lambda}{2}^2$ approximates mode area of a diffraction limited spot in free space, $A_{eff} = \frac{\left(\int_{-\infty}^{\infty} |E|^2 dy\right)^2}{\int_{-\infty}^{\infty} |E|^4 dy}$ is the effective mode area of a guided wave, and $E$ is the electric field (we use an effective mode area definition following Ref. [32]; we do not expect significant change in our conclusions with a use of more sophisticated definitions [33]). Fig. 1c plots the dispersion of $C_{opt}(\lambda)$ for waveguides made of TMDC materials and conventional semiconductors, including Si, GaAs and InP. Clearly, for any wavelength of interest, effective mode area sustained by TMDC waveguides is significantly smaller than that for traditional semiconductors. Strong optical field confinement, smaller mode area, and shorter effective wavelength suggest the potential of using TMDCs to create photonic components with significantly smaller device sizes and footprints.

Optical attenuation is another crucial factor determining performance of integrated nanophotonic systems. While a number of works examines in detail light-exciton coupling [21-24], little attention has been paid to sub-bandgap optical losses in bulk TMDC materials. Quantifying minute sub-bandgap optical losses, limited predominantly by crystalline quality and free-carrier absorption, necessitates highly sensitive measurements, such as photothermal deflection spectroscopy. Only few works since 1980s have performed such measurements for $MoS_2$ absorbance in the infrared range [27,28]. To infer optical attenuation of TMDC waveguides, in Fig. 1d we plot extinction coefficient, $= Im(\sqrt{\varepsilon})$, where $\varepsilon$ denotes respective material electric permittivity, for $MoS_2$, and Si, GaAs, and InP in the near infrared wavelength range ($850 nm < \lambda < 1200 nm$) based on ellipsometry and photothermal deflection spectroscopy measurements [27, 29, 34, 35]. (Note that we use extinction coefficient to compare intrinsic material properties, as actual waveguide losses depend also on fabrication imperfection and vary vastly across literature even for well-studied materials such as Si). Optical extinction of $MoS_2$ plateaus in sub-bandgap wavelength range ($\lambda \geq 1050\ nm$), which signifies the role of optically active defects [28]. The reported defect density for $MoS_2$ ($N \simeq 3.04 \times 10^{18} cm^{-3}$) [28] is 2 to 3 orders of magnitude higher than that of conventional optoelectronic semiconductors, implying stronger optical extinction in $MoS_2$ components. However, it is reasonable to expect that with further optimization of growth and fabrication techniques, defect density of $MoS_2$ and other TMDC materials may be significantly reduced. Assuming a simplistic model of light absorbance in sub-bandgap region being proportional to density of optically active defects [28], $\alpha = \sigma N$, we estimate to the first order approximation light absorbance for higher quality materials (here $\alpha = \frac{4\pi\kappa}{\lambda}$ is the optical absorption coefficient, $\sigma$ is the absorption cross-section for a single defect site, and $N$ is the density of defect sites). Hence, for $MoS_2$ with density of defects as low as $N \simeq 10^{15}\ cm^{-3}$ (i.e., comparable to that of amorphous Si [36]), reduction of optical extinction by 3 orders of magnitude is expected (see also Fig. 1e). In this case, $MoS_2$ waveguide insertion loss, $\gamma = \frac{20}{\ln(10)} Im(k)$, can go down to $\gamma \simeq 2.4\ dB/m$, which is comparable to low loss $Si_3N_4$ waveguides [37], while offering an unprecedented optical field confinement for dielectrics (here $Im(k) \simeq \frac{\omega \varepsilon_0 Im(\varepsilon) \int_0^h |E|^2 dy}{4 \int_{-\infty}^{\infty} S_z dy}$, $S_z = \frac{1}{2} Re[\mathbf{E} \times \mathbf{H}^*]\hat{z}$ is the energy flux density in the direction of propagation, $\hat{z}$).

To further compare $MoS_2$ with Si, GaAs and InP – semiconductors widely used in today's optoelectronics and integrated photonics – in Fig. 1e we study effective propagation length, $L_{eff}$, and optical confinement, $C_{opt}$, across sub-bandgap range (i.e., $\lambda \in [\lambda_{BG}, 1200\ nm]$, 1200nm cutoff is chosen due to availability of optical absorbance data for $MoS_2$). We define effective propagation length as $L_{eff} = \frac{Re(k)}{Im(k)}$, which indicates the number of effective wavelengths in the waveguide ($\lambda_{eff}$) per attenuation length. Clearly, $MoS_2$ waveguides exhibit much stronger confinement when compared to conventional semiconductor counterparts (~30% larger), while offering a comparable effective propagation length. This analysis suggests that $MoS_2$ is well suited for integrated photonics. Although we are not familiar with detailed studies of optical losses in sub-band gap regime for other TMDC materials (i.e., $MoSe_2$, $WSe_2$, $WS_2$), we expect that our analysis of losses and optical confinement pertains to these materials as well.

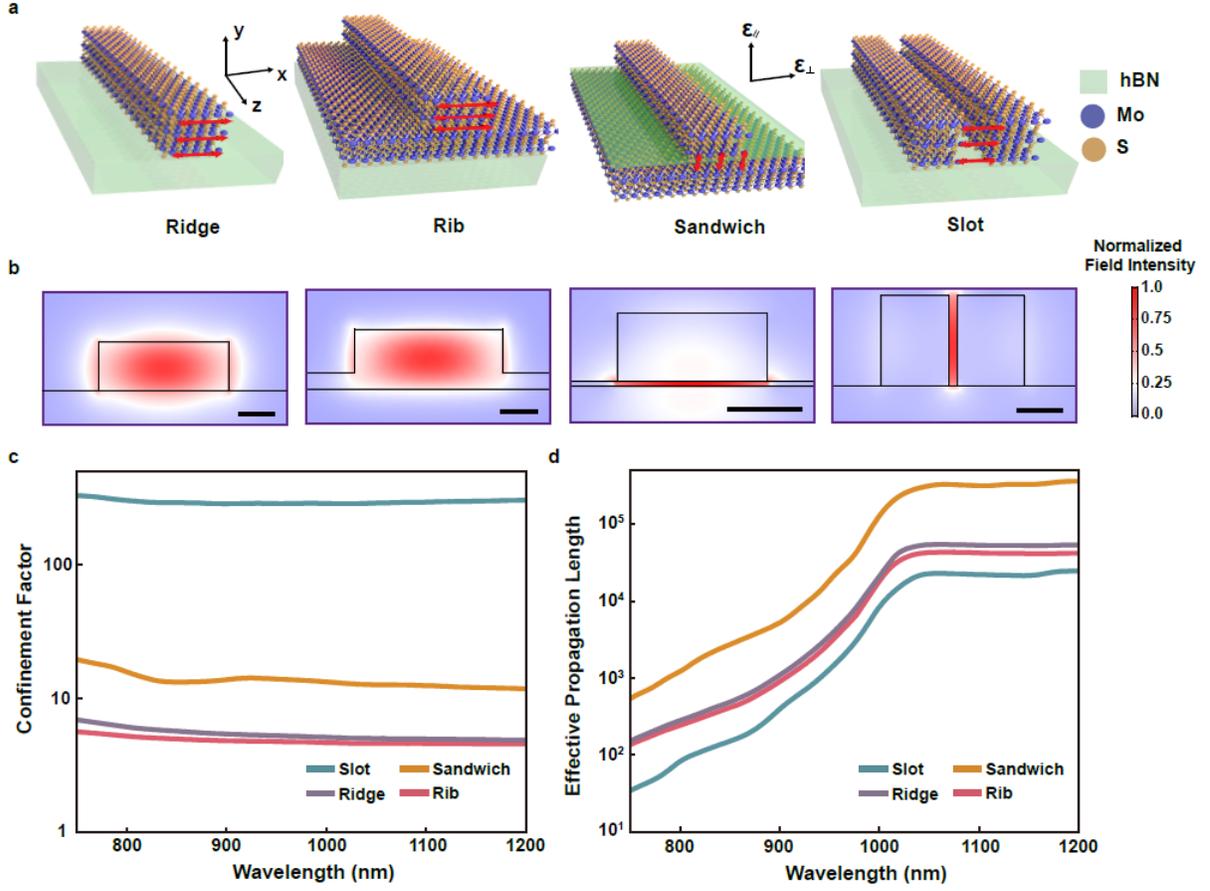

FIG. 2. MoS$_2$ waveguides for integrated van der Waals nanophotonics. (a) Schematic illustration of typical waveguide geometries, including ridge, rib, sandwich and slot waveguides made of MoS$_2$ and hBN. Dominant component of guided mode electric field polarization is indicated. (b) Calculated field intensity distribution of corresponding guided mode profiles. Scale bar is 100 nm. (c) Optical confinement and (d) effective propagation length of light guided four different waveguide configurations in a range 750 nm to 1200 nm.

Next, we study more complex 2D and 3D devices. For the sake of concept demonstration, we only consider MoS$_2$ – one of the well-studied TMDCs; our analysis can be readily transplanted to other TMDC materials of interest. In Fig. 2a, we consider four different waveguide configurations made of bulk MoS$_2$: ridge, rib, sandwich and slot waveguides. We use hexagonal boron nitride (hBN) – another van der Waals material – as a low index ($n \simeq 2$ in near-infrared spectrum [38]) cladding and spacer. Corresponding eigen-mode intensity profiles, $|E|^2$, are shown in Fig. 2b. In this analysis, due to a complex vectorial field configuration, we take the full tensorial nature of optical constants into account [19-21]. Fig. 2c shows wavelength dispersion of optimal optical confinement [32], $C_{opt}(\lambda) = \max A_o / A_{eff}$ ($A_o = \frac{\lambda^2}{4}$ and $A_{eff} = \frac{(\iint_{-\infty}^{\infty} |E|^2 \, dxdy)^2}{\iint_{-\infty}^{\infty} |E|^4 \, dxdy}$), which is found by performing a parametric sweep of waveguide dimensions to yield smallest effective mode area at each wavelength). Ridge and rib waveguides behave similarly to a planar 1D waveguide (Fig. 1c), as expected, since light is guided in MoS$_2$ core and is predominantly TE polarized as shown in Fig. 2a. Importantly, ridge and rib waveguides made of MoS$_2$ exhibit a significantly reduced footprint when compared to Si: on average ~30% smaller while offering >40% stronger optical field confinement. For example, with thickness $h = 120$ nm and wavelength $\lambda = 1200\ nm$, ridge waveguide made of Si requires a footprint of $w = 500$ nm to reach the best optical confinement. MoS$_2$ waveguide, on the other hand, is only 320 nm wide while maintaining a 50% higher confinement factor. Smaller footprint suggests a possibility of a denser photonic integration per single chip.

Slot and sandwich waveguides, in turn, allow for even further field confinement. In these structures light is squeezed and guided in narrow low index channels (air and hBN, respectively) allowing for a deeply subwavelength optical field localization. Interestingly, strong anisotropy of MoS$_2$ results in drastically different confinement factors for slot and sandwich waveguides. Hence, in a slot waveguide, mode is predominantly polarized along MoS$_2$ basal plane, implying that its dispersion is dictated by the large index contrast at interface between high-value in-plane refractive index component, $n_\perp \simeq 4.4$ and low index of air (n=1). The latter results in a very strong field localization within the air gap. On the contrary, for a sandwich waveguide, mode dispersion is determined by the index contrast between low-value out-of-plane component of MoS$_2$ refractive index tensor, $n_\parallel \simeq 3.3$, and hBN, refractive index of which is larger than air, leading to a weaker field confinement.

In Fig. 2d we plot effective propagation length, $L_{eff} = \mathrm{Re}(k)/\mathrm{Im}(k)$, where $\mathrm{Im}(k) = \frac{\omega\varepsilon_0 \int_0^w \int_0^h E^* Im(\bar{\bar{\varepsilon}}) E dxdy}{4 \iint_{-\infty}^{\infty} S_z dxdy}$, in which $Im(\bar{\bar{\varepsilon}})$ represents the tensor of imaginary permittivity of the material [21,26]. As expected, stronger field localization implies stronger light-MoS$_2$ interaction and therefore stronger optical attenuation, that is, smaller effective propagation length for ridge and rib waveguides. Slot waveguide exhibits even lower $L_{eff}$ (Fig. 2d) due to a much stronger optical field confinement (Fig. 2c). Sandwich waveguide, on the contrary, despite exhibiting smaller mode area (Fig. 2c), demonstrates more than six times the effective propagation length when compared to ridge and rib waveguides (Fig. 2d). We attribute this dynamics to mode polarization and anisotropy of MoS$_2$, as the field is predominantly polarized along the low loss out-of-plane direction (Fig. 2a). Deeply subwavelength optical field confinement in MoS$_2$ waveguides pertains to more complex 2D and 3D structures, as well as to other high-index TMDCs materials.

In addition to a reduced device footprint, we anticipate that optical structures made of bulk TMDCs possess a stronger light-materials interaction, which may lead to the enhancement of quantum and classical processes, including spontaneous light emission [39-44], nonlinear optical interactions [45,46], and high efficiency electro-optical switching [12,13]. Indeed, many photonic processes are inversely proportional to effective optical mode area $A_{eff}$ or mode volume $V_{eff}$. One of the most notable examples is the Purcell enhancement of the spontaneous emission of a quantum emitter placed in an optical cavity [44]: $F_P = \frac{3\lambda^3}{4\pi}\frac{Q}{V_{eff}}$, where $Q$ is a cavity quality factor.

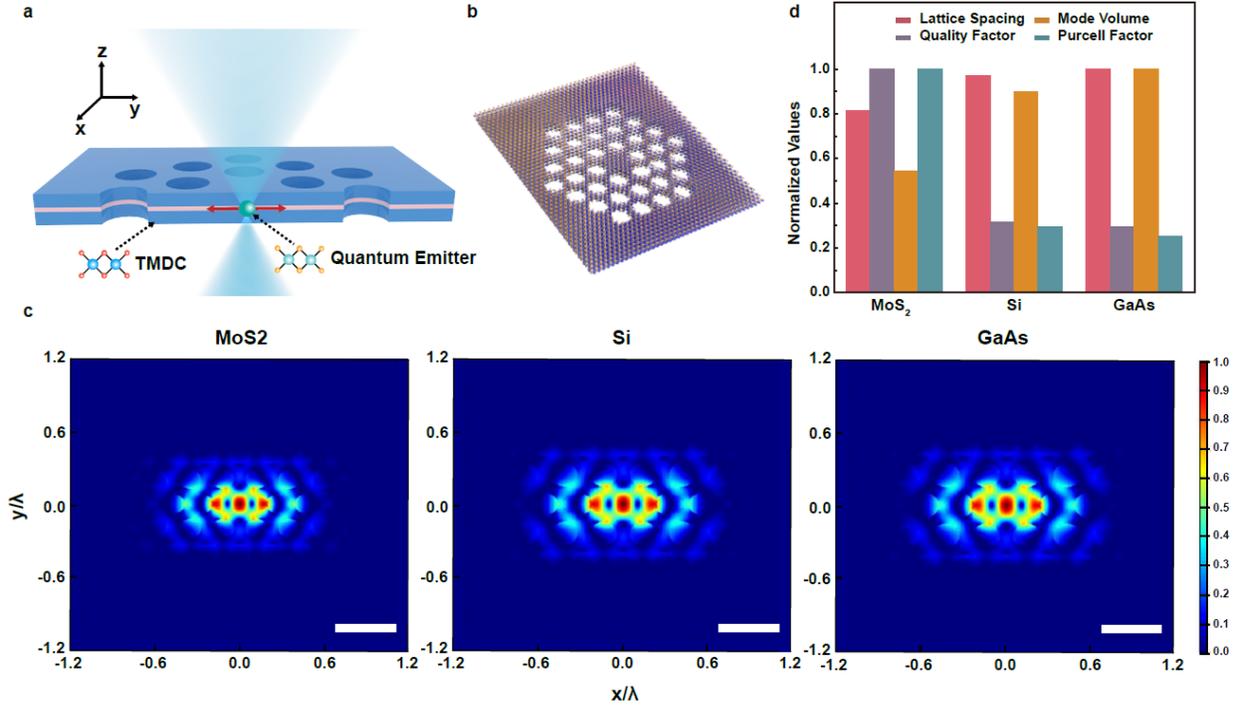

FIG.3. Spontaneous emission enhancement in MoS₂ photonic crystal cavity. (a) Schematic illustration of a point dipole emitter placed in a photonic crystal cavity made of TMDCs. (b) Schematic illustration of hexagonal lattice photonic crystal cavity made of MoS₂. (c) Normalized electric field intensity distribution of photonic crystal cavity modes at the resonance for MoS₂, Si and GaAs case, respectively. Scale bar is 500nm. (d) Performance comparison of photonic crystal cavities made from MoS₂, Si, and GaAs in terms of lattice spacing, Q factor, effective mode volume and Purcell enhancement factor.

To illustrate such a possibility, we consider spontaneous emission of a point emitter within a MoS₂ cavity, as shown in Fig. 3a. Specifically we consider MoS₂ nanocavity formed in a photonic crystal slab with a hexagonal lattice with a defect cavity formed in the middle [47] (Fig. 3b). We compare the results for MoS₂ photonic crystal cavity with corresponding Si and GaAs structures. In our design, considering lattice spacing, $a$, as a parameter, we fix hole radius, $r/a = 0.35$, and slab thickness, $d/a = 0.6$ and search for band gap conditions for TE-like modes (i.e., with electric field predominantly in-plane). We design structures with a cavity resonance at $\lambda = 1200nm$, i.e., below the bandgap for all three materials of interest (i.e., MoS₂, Si, and GaAs). To account for a complex field configuration and MoS₂ anisotropy, we derive a modified Purcell enhancement factor:

$$F_P \to \frac{3\varepsilon_0 \lambda^3}{2\pi^2} Q \frac{[\mathbf{n}\mathbf{E}(\mathbf{r}_0)][\mathbf{n}\mathbf{E}^*(\mathbf{r}_0)]}{\int [\varepsilon_0 \mathbf{E}\bar{\bar{\varepsilon}}\mathbf{E}^* + \mu_0 |\mathbf{H}|^2] dV}, \qquad (1)$$

where $\mathbf{n}$ and $\mathbf{r}_0$ denote dipole emitter orientation and position, respectively. For a y-oriented point source (as shown in Fig. 3a), we obtain $F_P = \frac{3\lambda^3}{2\pi^2 n_\perp^2} \frac{Q}{V_{eff}}$ with $V_{eff} = \frac{\int [\varepsilon_0 \mathbf{E}\bar{\bar{\varepsilon}}\mathbf{E}^* + \mu_0 |\mathbf{H}|^2] dV}{\varepsilon_0 \varepsilon_\perp |E_y(\mathbf{r}_0)|^2}$. In Fig.

3c we plot mode profiles at the resonance for all three cases (i.e., MoS$_2$, Si, and GaAs). Much smaller optical mode volume and physical footprint of a MoS$_2$ cavity as compared to Si and GaAs counterparts is clearly seen. Hence, the lattice spacing of MoS$_2$ photonic crystals is 20% smaller ($a \sim 286$ nm vs $a \sim 350$ nm) and effective mode volume is 45% smaller than that for Si and GaAs cavities. Fig. 3d summarizes these findings in a bar chart. Our analysis further shows a more than threefold higher quality factor, $Q$, for MoS$_2$ structures as compared to Si and GaAs. We attribute this dynamics to a smaller optical scattering cross section of the MoS$_2$ cavity owing to its smaller optical mode volume, and, thus, weaker coupling with the free space. Larger quality factor and smaller effective mode volume lead to a stronger Purcell enhancement of spontaneous emission, as expected (see Eq. (1)). We find that the cavities made of MoS$_2$ exhibit a Purcell enhancement four times higher than that for Si and GaAs cavities. Stronger Purcell enhancement is a direct indication of an enhanced light-material interaction in MoS$_2$ nanostructures. We anticipate a similar dynamics for other TMDC materials, suggesting their promise for quantum optics applications [43,44].

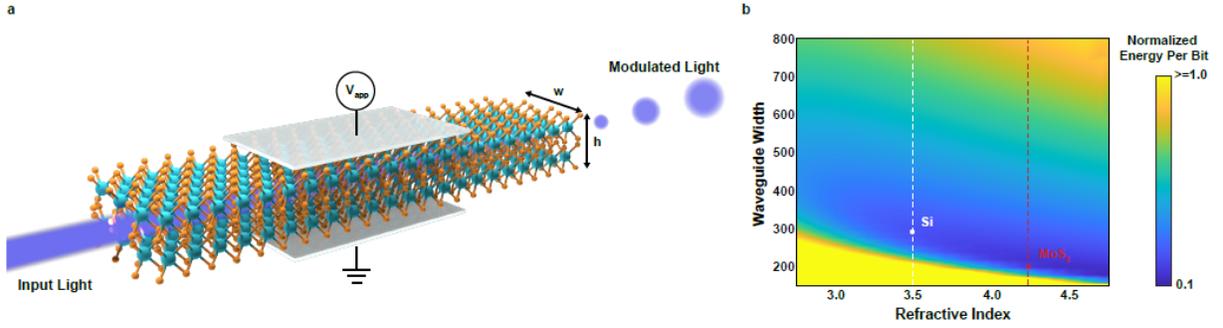

FIG. 4. TMDC electo-optical modulators. (a) Schematic illustration of waveguide electro-optic modulator made of MoS$_2$. (b) Normalized energy per bit for a travelling wave modulator with device width and refractive index variations. Here, waveguide thickness of 120 nm and operating wavelength of 1200 nm are assumed. Dashed lines indicate refractive indices of Si and MoS$_2$, respectively.

Finally, we show that reduced device footprint and enhanced optical field confinement (i.e., smaller mode area, $A_{eff}$) may translate onto higher energy efficiency of integrated electro-optical components. For this purpose, we consider a travelling wave electro-optical modulator [12,13] shown schematically in Fig. 4a. For the sake of simplicity, we assume that the modulator is a rectangular waveguide of length $L$, width $w$, and thickness $h$. Energy per bit of information for such a modulator may be estimated as [48,49] $\mathcal{E} = \frac{1}{2}CV_b^2$, where $C \propto \frac{Lw}{h}$ is the capacitance and $V_b = E_b h$ is the applied voltage, $E_b$ is the bias electric field. Under an applied bias material refractive index and/or extinction coefficient are perturbed, $\Delta\varepsilon = \chi E_b$, resulting in perturbation of the guided mode wavevector $\Delta k = \frac{\omega \varepsilon_0 \chi E_b \int_0^w \int_0^h |E|^2 dx dy}{4 \iint_{-\infty}^{\infty} S_z dx dy}$, where $\chi$ describes an effective modulation strength. Note that here we do not detail physical mechanism of a possible modulation, but rather provide a parametric study for a generic scenario (which may be based on an electro-optical effect, absorption modulation, or carrier injection). Modulator effective length may then be estimated as $L \simeq \frac{2\pi}{\Delta k}$, giving a normalized energy per bit:

$$\frac{\varepsilon\chi}{V_b} \propto \frac{8\pi \iint_{-\infty}^{\infty} S_z dxdy}{\omega\varepsilon_0 \int_0^w \int_0^h |E|^2 dxdy} w. \qquad (2)$$

Assuming a similar modulation strength, $\chi$, and applied voltage, $V_b$, for different materials, it is evident that switching energy scales with effective mode area and waveguide width, $w$, as expected [48]. Figure 4b shows a parametric analysis of the normalized energy per bit (Eq. (2)) as a function of modulator refractive index and waveguide width (here we assume $\lambda = 1200\ nm$ and $h = 120\ nm$ for consistency). Clearly, smaller waveguide footprint, i.e., smaller width, and higher refractive index yield smaller normalized switching energy. Specifically, comparing $MoS_2$ performance to Si we find that $MoS_2$ modulator offers nearly twice smaller footprint ($w = 280\ nm$ vs $w = 500\ nm$ for Si) and 42% higher optical confinement, while exhibiting 45% smaller normalized switching energy when compared to a Si modulator (again considering that $\chi$ and $V_b$ are equivalent for both materials). This analysis suggests that smaller footprint and more energy efficient active optical components may be designed with $MoS_2$ or other high index TMDC materials.

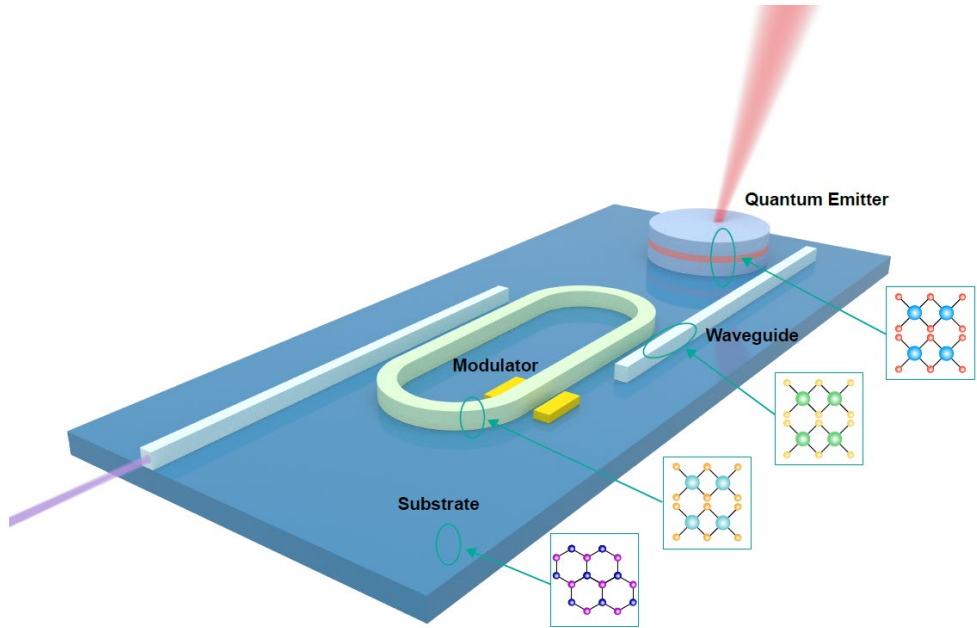

FIG. 5. All van der Waals integrated nanophotonics. An artistic illustration of a concept of an all van der Waals integrated nanophotonics, in which all functional components and waveguides are made of 2D materials. Light is emitted, detected and manipulated within bulk 2D semiconductors and heterostructures.

In summary, we have surveyed several examples of classical and quantum nanophotonic structures made of bulk TMDC materials. We showed that such devices may offer smaller footprint, higher energy efficiency and stronger light-materials interaction, as compared to semiconductors traditionally used in integrated photonics. With further perfection of growth and fabrication techniques [50,51] and developments in materials discovery and synthesis [52], wider class and higher quality TMDC materials are expected. Enhanced performance of TMDC optical components suggests they may complement, or perhaps, even replace conventional semiconductors in integrated photonics circuits. Notably, self-passivated surfaces free of dangling bonds [1-4], high electron mobility in nanometer thick films [1-3], lattice match-free van der Waals bonding [1,4], and compatibility with foundry processes [21,22,24,43] suggest that different TMDCs materials may be integrated with diverse material platforms. We specifically, envisage an all van der Waals integrated nanophotonics (Fig. 5) where different TMDC materials are brought together to create functional optical devices and systems where light is guided and processed within van der Waals materials with efficiency and performance exceeding covalent integrated optics in use today.

**Acknowledgements**

We thank UCLA Samueli for startup support and UCLA Council on Research Faculty Research Grant. We thank Aaswath Raman and his group for access and support with Lumerical.